\documentclass{PoS}

\usepackage{url}
\newcommand{\be}{\begin{equation}}
\newcommand{\ee}{\end{equation}}
\newcommand{\bea}{\begin{eqnarray}}
\newcommand{\eea}{\end{eqnarray}}
\newcommand{\bi}{\begin{itemize}}
\newcommand{\ei}{\end{itemize}}
\newcommand{\ben}{\begin{enumerate}}
\newcommand{\een}{\end{enumerate}}

\newcommand{\lc}{\left[}
\newcommand{\rc}{\right]}

\title{Scale-Invariant Resonance Tagging \\in Multijet Final States}

\ShortTitle{Scale-Invariant Resonance Tagging}

\author{\speaker{Juan Rojo}\thanks{This work is supported by a Marie Curie 
Intra--European Fellowship of the European Community's 7th Framework Programme under contract number PIEF-GA-2010-272515. }\\
        PH Department, TH Unit, CERN, CH-1211 Geneva 23, Switzerland\\
        E-mail: \email{juan.rojo@cern.ch}}

\abstract{
In this contribution we study the resonant pair production 
of heavy particles in hadronic
final states using jet substructure techniques.
We discuss a recently proposed resonance tagging strategy, which interpolates
between the highly boosted and fully resolved regimes, leading to
uniform signal efficiencies and background rejection
rates for a broad range of  masses.
With this method, one can efficiently replace independent experimental searches, based on different
final state topologies, with a single common analysis.
We show using this strategy that the LHC has sensitivity
to the enhanced resonant production of
 Higgs boson pairs decaying into $b\bar{b}$ pairs in generic
New Physics scenarios.
}

\FullConference{XXI International Workshop on Deep-Inelastic Scattering and Related Subjects -- DIS2013,\\
		22-26 April 2013\\
		Marseille, France}

\begin{document}

\paragraph{Scale-invariant resonance tagging.}

Searches for new physics in multijet events
are an important ingredient of the LHC physics program.
A  challenge in such searches for new phenomena is the overwhelming QCD multijet 
background.
A range of techniques becomes necessary in order
to identify  important categories of
jets~\cite{Salam:2009jx}, and thus improve the QCD background rejection.
In particular, jet substructure 
tools have been the subject of substantial development in the last
years~\cite{Abdesselam:2010pt}, and allow to improve the signal over background
ratio of an important number of boosted and semi-boosted final states.

\begin{figure}[h]
\centering
\vspace{-10mm}
\includegraphics[scale=0.33]{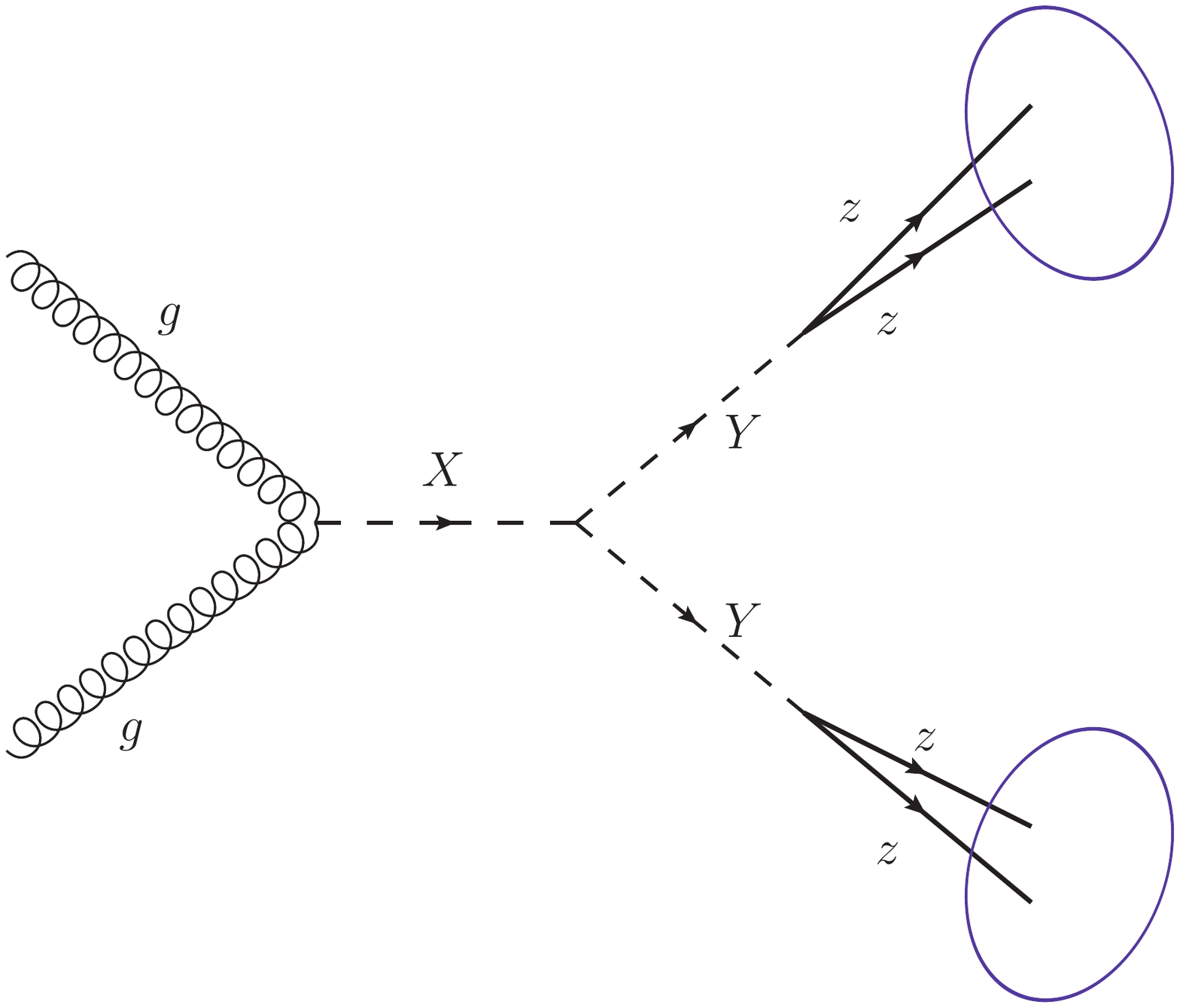}
\includegraphics[scale=0.35]{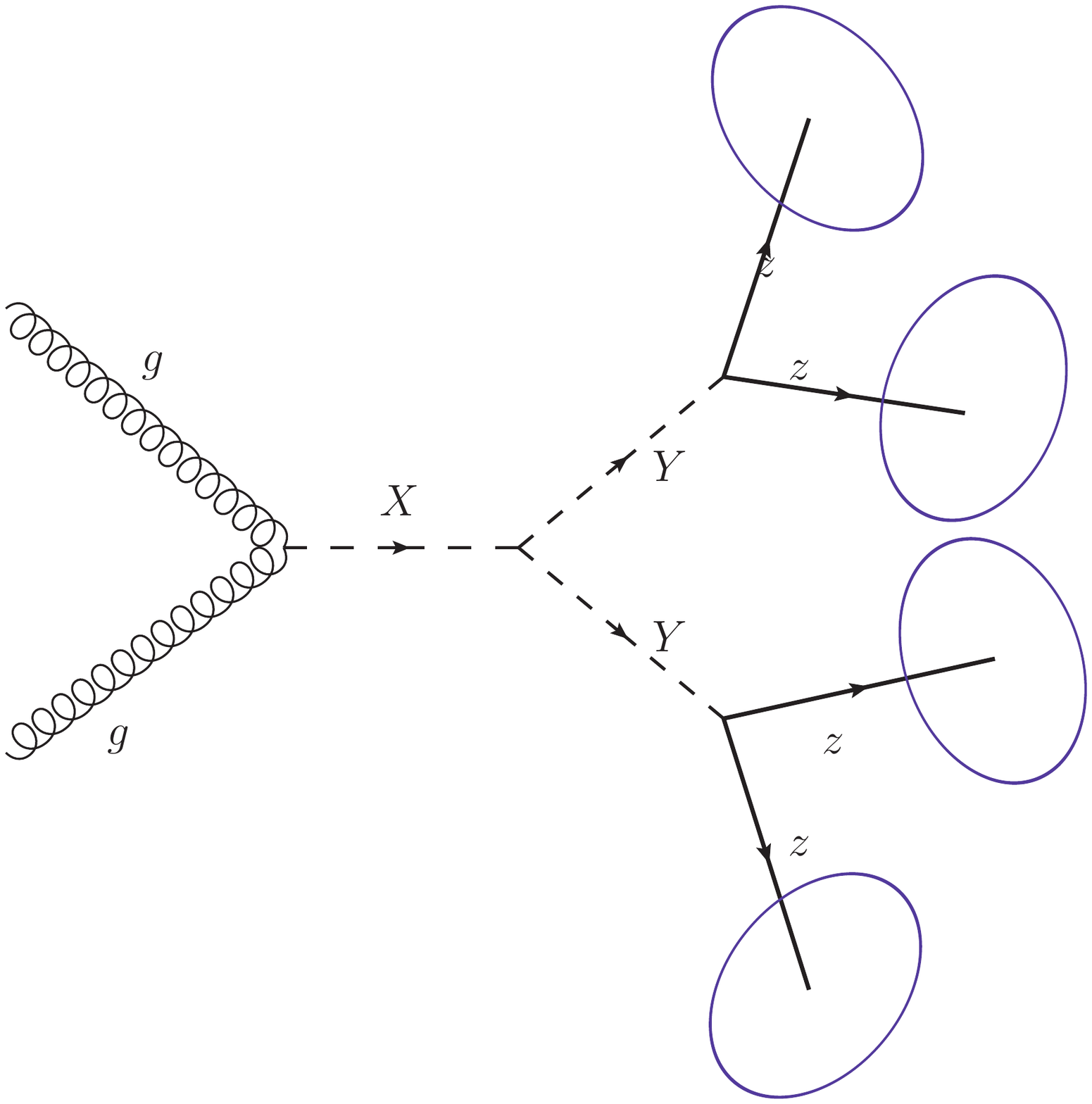}
\vspace{-27mm}
\caption{\small   Schematic diagrams
for the generic process $pp \to X \to 2 Y \to 4 z$ 
in the boosted (left plot) regime, corresponding
to large values of the mass ratio 
$r_M=M_X/2M_Y$, and in the resolved (right plot) regime,
corresponding to small values of $r_M$.
}
\label{fig:toy}
\end{figure}

In an important class of new physics  models, paired production of
resonances dominates. These processes are
generically of the form
$pp\to X \to 2Y \to 4{~\rm partons}$, with $X$ and $Y$ being
 heavy particles, as illustrated schematically in Fig.~\ref{fig:toy}.
The mediator $X$ of this production 
could be an
exotic particle from a new strongly coupled sector,
or a resonance from extra dimensions.
The
$Y$ resonance could be either some BSM particle or on the other hand
some SM particle ($W,Z$ or Higgs)
that subsequently decays into quarks and gluons, observed as
jets in the LHC detectors.

The generic four-parton processes
lead to  distinct final state signatures depending on the interplay
between the masses of the two intermediate resonances, $M_X$ and
$M_Y$, which is characterized by the dimensionless variable
$r_M\equiv M_X/2M_Y$.
For a large mass ratio, $M_X\gg M_Y$, the $Y$ resonances
 will be produced  boosted, and typically the decay products
of each of the two $Y$ resonances will be collimated into a single {\it fat}
jet. 
On the other hand, for $M_X\sim 2 M_Y$, the $Y$ resonances
will be produced nearly at rest, decaying into four well
separated jets. Existing searches assume either the highly
boosted or fully resolved regimes, and by doing so exclude
a potentially large region of the new physics parameter space. 

In Ref.~\cite{Gouzevitch:2013qca} we introduced 
a jet reconstruction
and analysis strategy that
can be applied simultaneously to the boosted and resolved regimes, and
that provides a smooth interpolation between them.
This was  achieved by merging the boosted-regime strategies, based
on the BDRS mass-drop tagger~\cite{Butterworth:2008iy}, with a suitable  strategy for the resolved 
four-jet regime, in our case dijet mass pairings.
A set of quality requirements on the $p_T$ and angular separation of
jets and subjets was imposed to ensure a roughly constant tagging efficiency and a smooth
interpolation between the two limiting regimes.
This strategy has the potential to make the experimental searches
for this particular topology more
efficient and  to allow a wider range of new physics scenarios to be
explored within a single analysis.

\begin{figure}[h]
\centering
\includegraphics[scale=0.37]{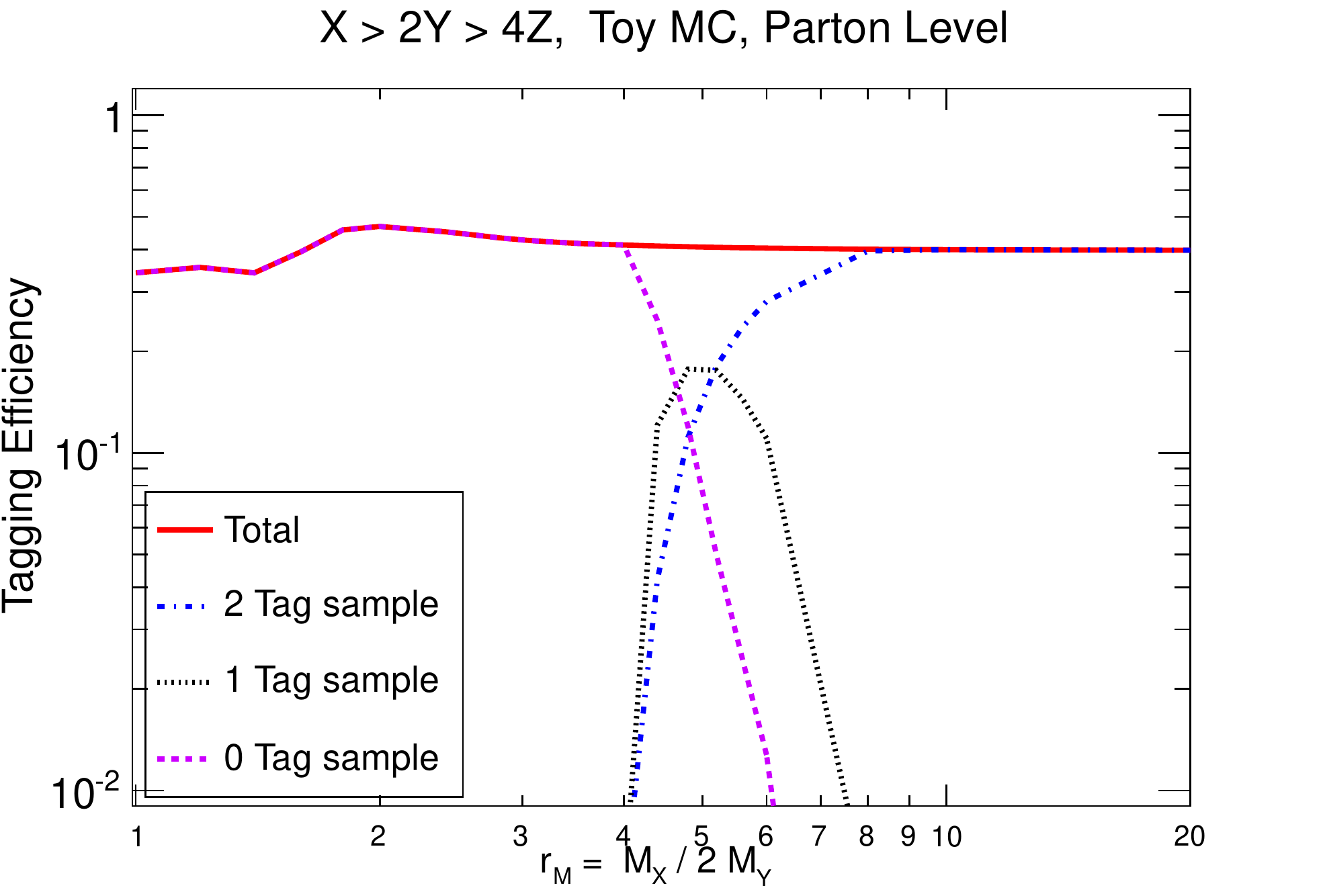}
\includegraphics[scale=0.37]{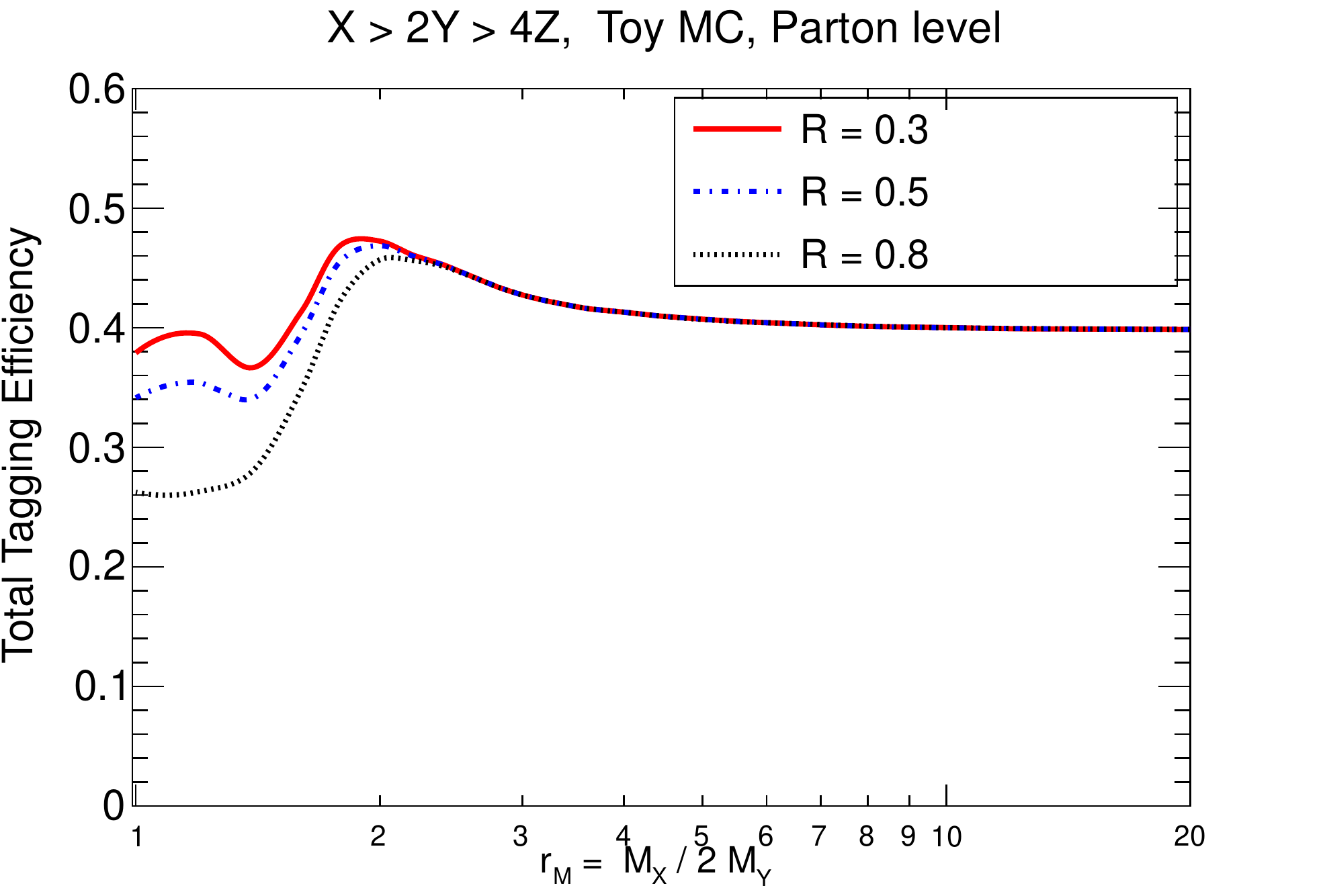}
\includegraphics[scale=0.37]{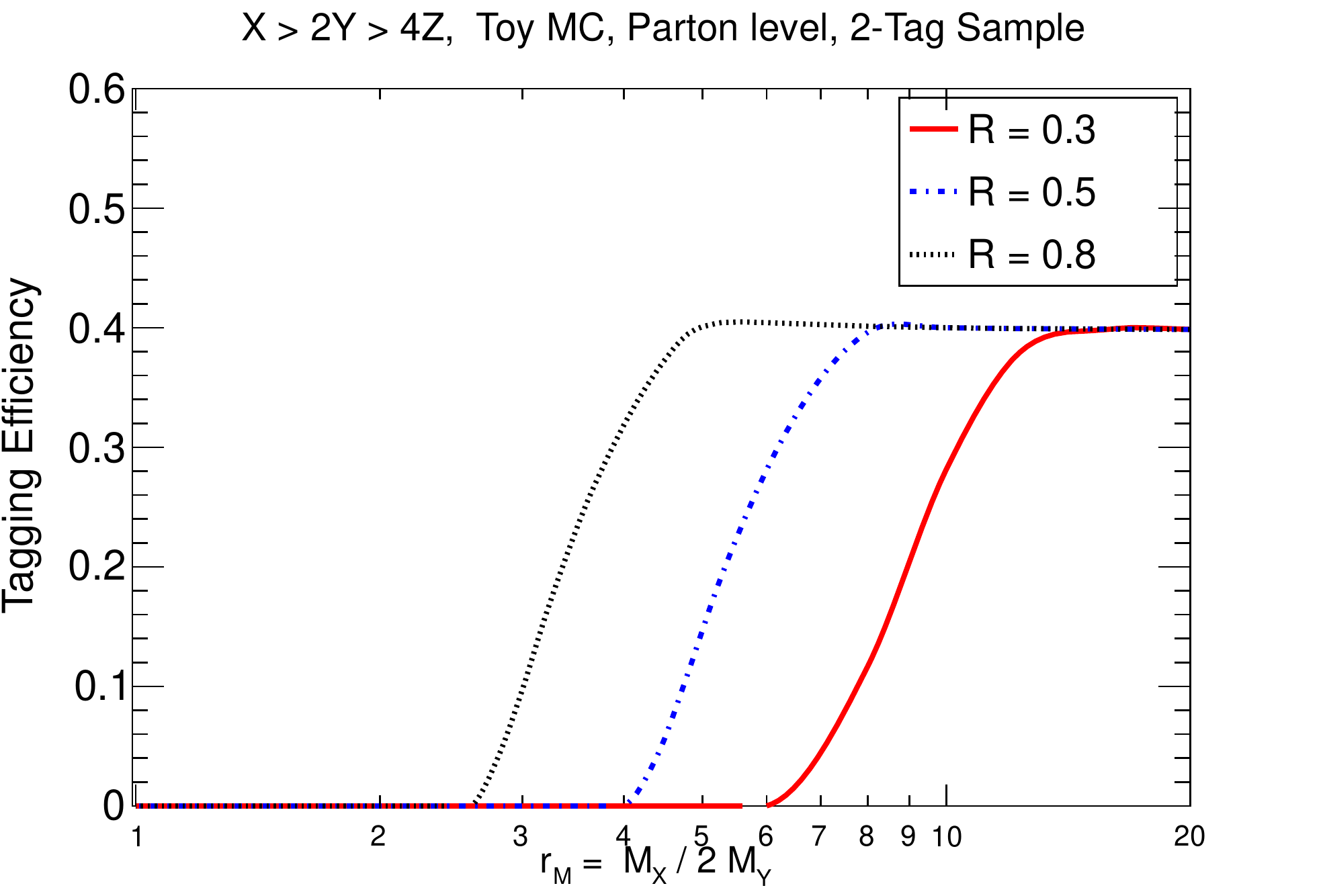}
\includegraphics[scale=0.37]{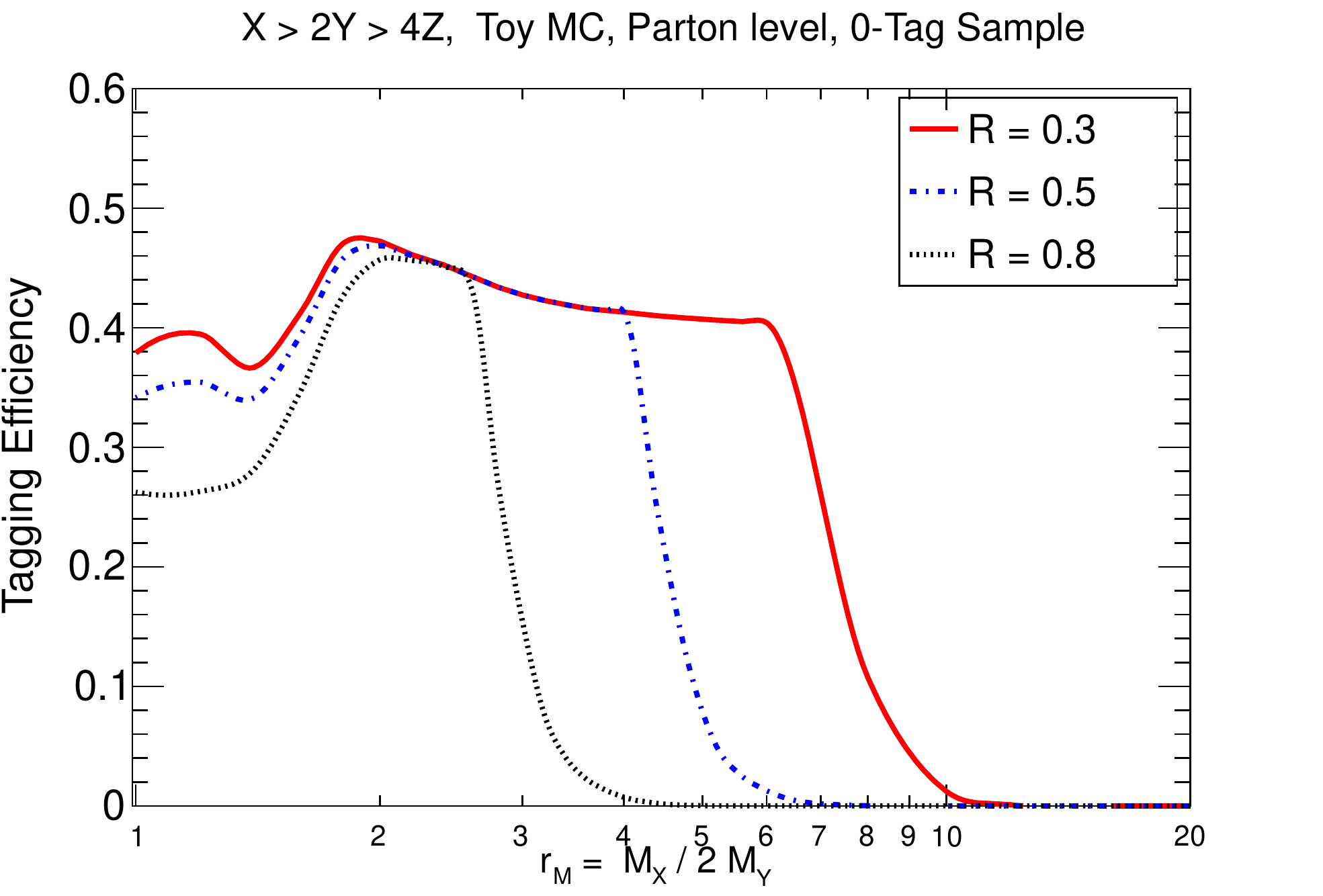}
\caption{\small  Upper left plot: The efficiency of the resonance
pair tagging algorithm as a function of resonances mass ratio
for parton-level toy Monte Carlo events.
We show both the total efficiency and the break-up for the
boosted, intermediate and resolved categories.
Upper right plot: the total tagging efficiency for different values
of the jet radius $R$.
Lower plots: the tagging efficiencies for the boosted category
(left plot) and for the resolved category (right plot), for different values of the
jet radius $R$.
}
\label{fig:efficiency-parton}
\end{figure}

The method sketched above was implemented into a code 
based on {\tt FastJetv3}~\cite{Cacciari:2011ma}.
We used the anti-$k_T$ jet clustering algorithm with $R=0.5$, though as
discussed below the dependence of our results with $R$ is very mild.
In Fig.~\ref{fig:efficiency-parton} we show 
the efficiency of the resonance
pair tagging algorithm as a function of $r_M$
for  events for the
$pp\to X \to 2Y \to 4z$ topology.
Parton level events were
produced with an in-house toy MC event generator, and then
showered with {\tt Pythia8} to produce realistic hadron level events.
In  Fig.~\ref{fig:efficiency-parton}  we show that the total tagging efficiency at parton level
 is approximately constant
over most of the $r_M$ range, with the boosted topology  (2-tag sample) dominating
at large $r_M$, the resolved topology dominating at small $r_M$ (0-tag sample), with
the 1-tag topology providing a smooth interpolation in the intermediate $r_M$ region. 
This approximate constant tagging efficiency also holds true at hadron level.

We also show in  Fig.~\ref{fig:efficiency-parton}   that the  tagging efficiency is
independent of the value of the jet radius $R$ used, except close
to  production threshold.
Such a stability with respect to $R$ is an useful feature of an
analysis technique, since it allows to adjust the value of $R$ 
to the specific process under consideration\footnote{For instance,
in the case of a heavy resonance decaying into jets, a large value of
$R$ is advantageous to improve mass resolution since it collects
most of the final state radiation.~\cite{Cacciari:2008gd}}, while keeping constant the overall
efficiencies for all values of the candidate resonance mass.
As shown in the lower plots of Fig.~\ref{fig:efficiency-parton}, 
the relative fraction of boosted over resolved
topologies depends strongly in the value of jet radius $R$, but thanks to
the quality requirements adopted in our resonance tagging strategy, their sum turns out to be
$R$-independent to a  good approximation.

\paragraph{Background rejection.}

For multijet topologies with at most four leading jets,
but without high $p_T$ leptons or
missing $E_T$, the dominant Standard Model
 background  is QCD jet production.
There are several ways in which QCD radiation can mimic the
conditions for resonance tagging, for instance, fake mass drops
can be generated with sufficiently symmetric splitting of a quark or gluon.

We define the  dijet cross section, for each
value $M$ of the candidate resonance mass, as the number
of QCD  events that survive the
 basic selection cuts and lead to
 an invariant mass
within the mass resolution window around $M$ given by
$\lc M(1-f_m),M(1+f_m)\rc$, and with the
the two leading jets are separated in rapidity by less than
$\Delta y_{\rm max}$. 
After resonance tagging was applied to the QCD multijet events, 
$b$-tagging was also required, with the condition that at least one
$b$-quark should be identified within each of the two Higgs candidate jets.
The settings for $b$-tagging were taken to be similar to the default ones
in ATLAS and CMS.
QCD multijet  background events has been generated with {\tt Pythia8}~\cite{Sjostrand:2007gs}.

\begin{figure}[h]
\centering
\includegraphics[scale=0.37]{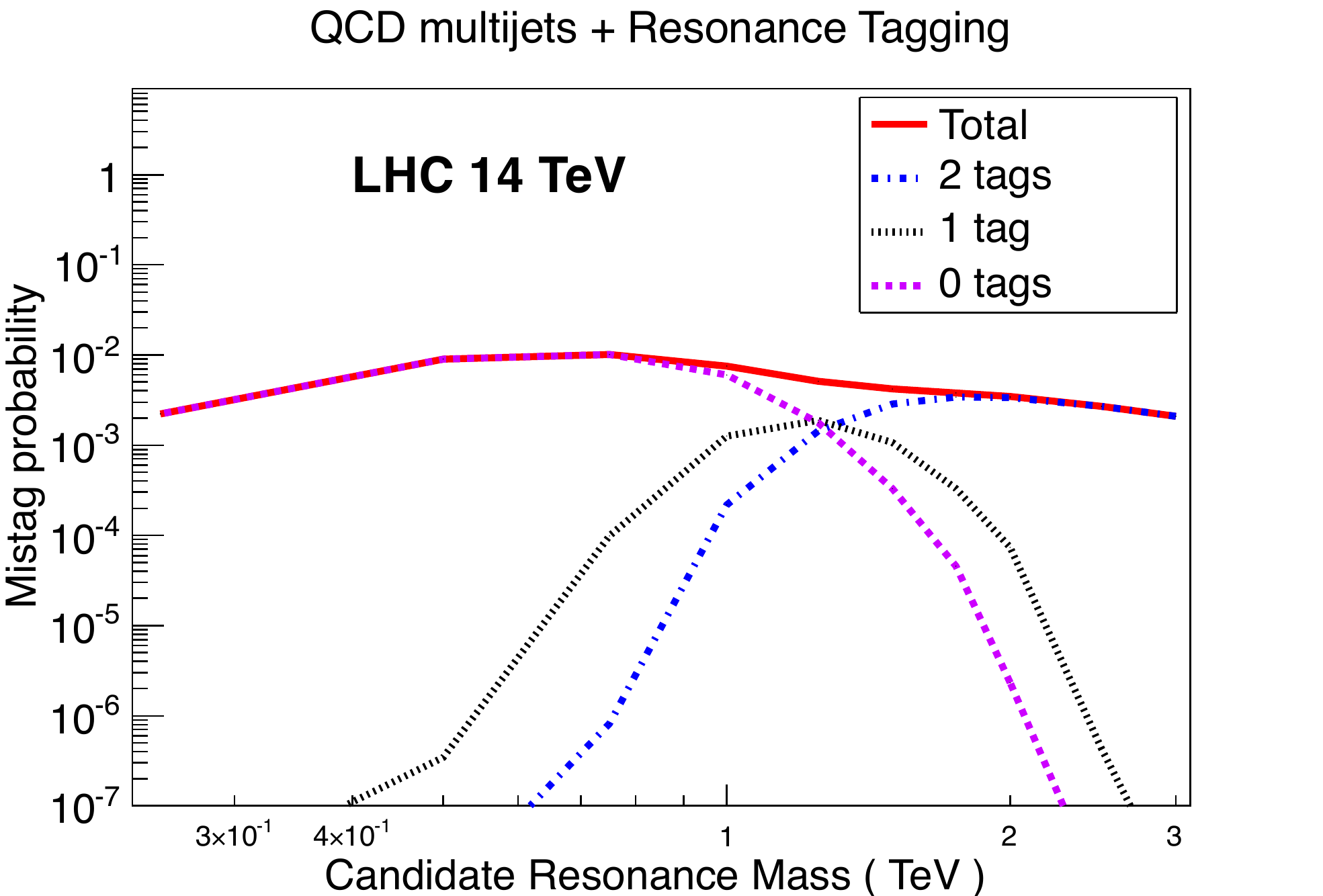}
\includegraphics[scale=0.37]{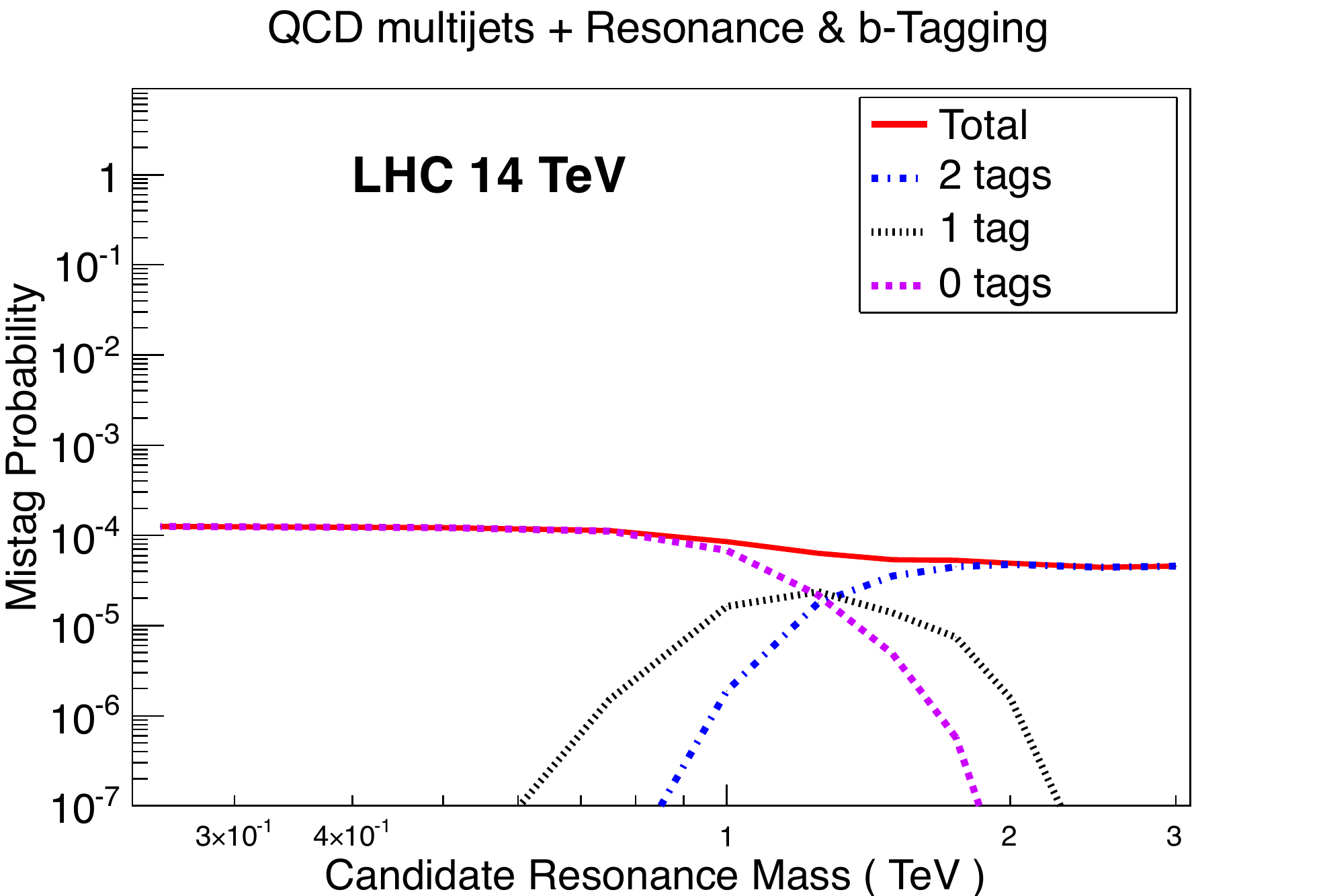}
\caption{\small The mistag probability of QCD multijet events, after resonance tagging (left plot) and after resonance and
$b$-tagging (right plot).
We show both the total mistag rate and its decomposition into the three
possible categories: 0-tag (where QCD multijets mimic the signal resolved
topology), 1-tag and 2-tags (where now the background events mimic the
boosted topology by means of fake mass drop tags).
The total mistag probability is reasonably independent of the candidate
resonance mass.
}
\label{fig:mistagprob}
\end{figure}

In Fig.~\ref{fig:mistagprob} we show 
the background rejection factors,
defined as
 the fraction of the QCD dijet events which
are mistagged as arising from a heavy resonance,
both with and without $b$-tagging.
We show  both the total mistag probability and the division
in 2-tag, 1-tag and 0-tag cathegories.
Is clear that
 the background rejection probability is approximately scale invariant,
thanks to the consistent contributon of each of the cathegories:
the 0-tag cathegory dominates at small $M$, while the 2-tag
cathegory dominates at large $M$.

\paragraph{New physics in the $2H \to 4b$ final state.}

As a phenomenologically relevant application of the scale-invariant resonance tagging technique, 
the method was applied in Ref.~\cite{Gouzevitch:2013qca}
to study the discovery potential of resonant Higgs pair production in the
$4b$ final state at the LHC.
We derived first of all model-independent bounds to generic
BSM scenarios that lead to enhanced resonant Higgs pair production, and then
interpreted these bounds in terms of Radion and Graviton production
in the context of warped extra dimensions models~\cite{Randall:1999vf}.
The theory predictions for Radion and Graviton production were
obtained from {\tt MadGraph5}~\cite{Alwall:2007st} interfaced
to {\tt Pythia8} for the parton shower.

\begin{figure}[h]
\centering
\includegraphics[scale=0.37]{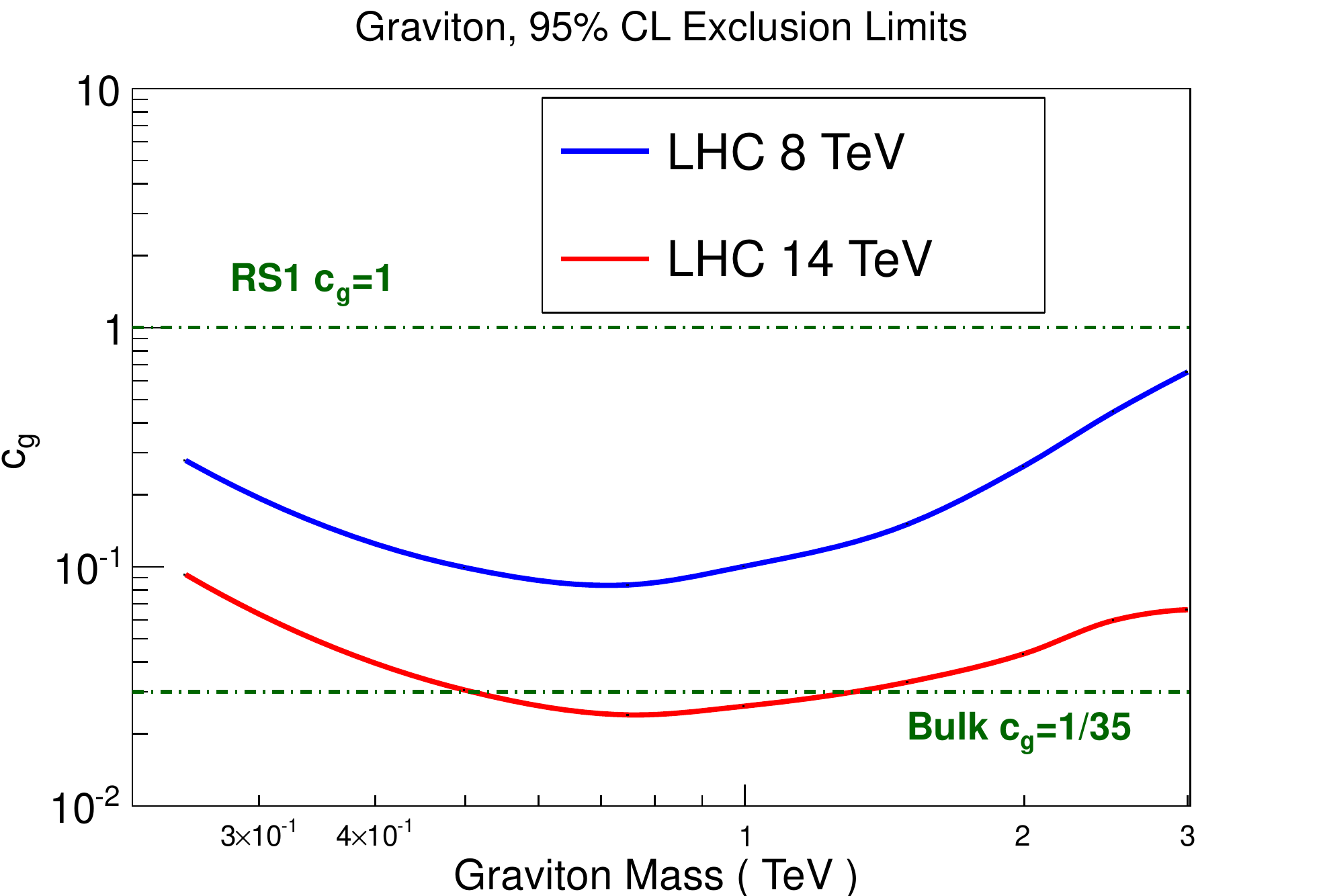}
\includegraphics[scale=0.37]{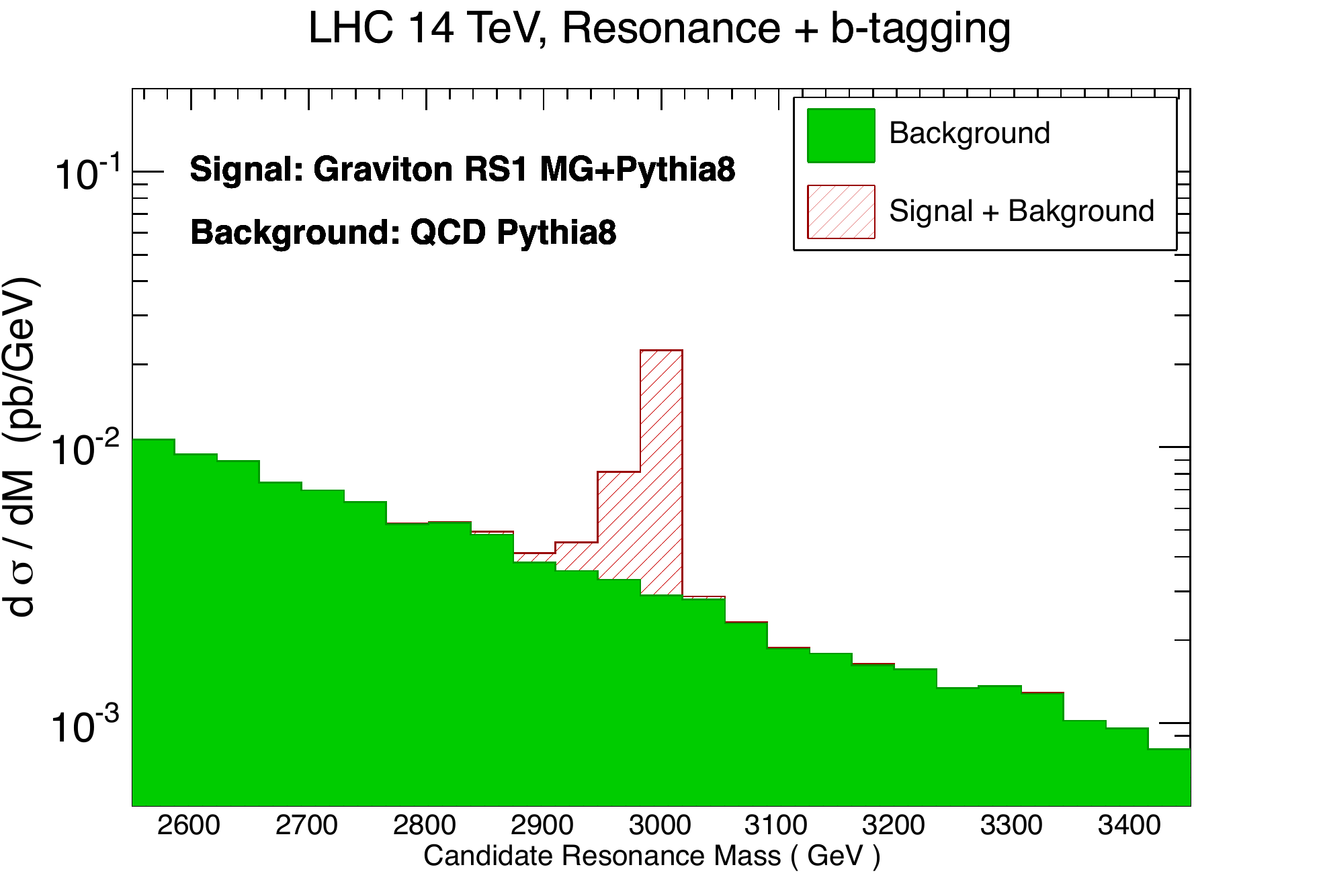}
\caption{\small Left plot: the values of the massive KK graviton-gluon
coupling $c_g$ that can be excluded at the 95\% CL as a function
of the graviton mass at the LHC 8 and 14 TeV.
Right plot: the signal plus background dijet tagged cross section, for
the case of a $M_g=3$ TeV KK Graviton production, compared to the prediction from
QCD multijet production only, after resonance and $b$-tagging have been applied
both to signal and background events. }
\label{fig:excluded}
\end{figure}

In Fig.~\ref{fig:excluded} we show the values of the 
massive KK graviton-gluon
coupling $c_g$ that can be excluded at the 95\% CL as a function
of the graviton mass for 8 and 14 TeV using the scale-invariant 
resonance tagging in the $4b$ final state.
We find that most of the relevant parameter space for radion and graviton
production in warped extra-dimensions scenarios
can be excluded using the $4b$ final state, and our technique allows
to cover most of the relevant Graviton mass range. 
Therefore, using our technique the $4b$ final states becomes a complementary handle
to the searches for similar BSM phenomena in other 
final states.
In Fig.~\ref{fig:excluded} we also show the signal plus background dijet b-tagged cross section
corresponding for
the case of  $M_g=3$ TeV Graviton production, compared to the predictions from
QCD multijet production only.
A clear signal mass peak arises on top of the monotonically decreasing QCD background. 

Applications of the scale-invariant resonance tagging idea
to other relevant problems include merging the threshold and boosted
searches in fully hadronic top quark production and the search for
enhanced Higgs pair production in other production channels such as vector-boson-fusion.

\providecommand{\href}[2]{#2}\begingroup\raggedright\endgroup

\end{document}